\documentclass[aps,prd,10pt,onecolumn,groupedaddress,nofootinbib,preprintnumbers,superscriptaddress]{revtex4-2}  

\usepackage{graphicx,mathtools,tabu,array}
\usepackage{epstopdf}
\usepackage{amsmath}
\usepackage{amsfonts}
\usepackage[colorlinks=true,citecolor=red,linkcolor=blue,breaklinks=true]{hyperref}
\usepackage{accents}
\usepackage[figure, table]{hypcap}
\usepackage{amssymb}
\usepackage{appendix}
\usepackage{comment}
\usepackage{bbold}
\usepackage{xcolor}
\usepackage{slashed}
\usepackage{subfigure}
\usepackage{setspace}
\usepackage{footnote}
\usepackage{multirow}
\usepackage{braket}
\usepackage[normalem]{ulem}
\usepackage[utf8]{inputenc}
\usepackage{mathrsfs}
\usepackage{longtable}
\usepackage{enumitem}
\usepackage{booktabs} 
\usepackage{subfigure}

\begin{document}

\preprint{IFT-UAM/CSIC-26-69}

\title{{\tt BlackHawk v3.0}:  Hawking Radiation from Regular Black Holes}

\author{Alexandre Arbey}
\email{alexandre.arbey@ens-lyon.fr}
\affiliation{Université Lyon 1, CNRS, IP2I, UMR 5822, Villeurbanne, France}

\author{Marco Calz\`a}
\email{marco.calza@unitn.it}
\affiliation{Department of Physics, University of Trento, Via Sommarive 14, 38123 Povo (TN), Italy}
\affiliation{Trento Institute for Fundamental Physics and Applications (TIFPA)-INFN. Via Sommarive 14, 38123 Povo (TN), Italy}

\author{L\'ea Malacher}
\email{lmalacher@ip2i.in2p3.fr}
\affiliation{Université Lyon 1, CNRS, IP2I, UMR 5822, Villeurbanne, France}

\author{Davide Pedrotti}
\email{davide.pedrotti-1@unitn.it}
\affiliation{Department of Physics, University of Trento, Via Sommarive 14, 38123 Povo (TN), Italy}
\affiliation{Trento Institute for Fundamental Physics and Applications (TIFPA)-INFN. Via Sommarive 14, 38123 Povo (TN), Italy}

\author{Yuber F. Perez-Gonzalez}
\email{yuber.perez@uam.es}
\affiliation{Departamento de F\'{i}sica Te\'{o}rica and Instituto de F\'{i}sica Te\'{o}rica (IFT) UAM/CSIC, Universidad Aut\'{o}noma de Madrid, Cantoblanco, 28049 Madrid, Spain}

\begin{abstract}
We present \texttt{BlackHawk v3.0}, a major update of the public code designed to compute Hawking radiation spectra of black holes. Building upon previous versions, this release considerably extends the range of black hole geometries that can be studied by implementing several new spherically symmetric metrics: the Bardeen and Hayward regular black holes, the Simpson-Visser and Peltola-Kunstatter black-bounces, the D'Ambrosio-Rovelli black hole-to-white hole metric, and the Babichev-Charmousis-Lehébel black hole. For each metric, we compute the corresponding Hawking temperatures and greybody factors, enabling the determination of primary Hawking emission spectra for particles of different spins. The greybody factors are obtained through dedicated numerical routines based on the companion code \texttt{GrayHawk}. Additionally, \texttt{BlackHawk v3.0} introduces several technical improvements aimed at enhancing precision and efficiency, providing a highly versatile tool. The code is publicly available at https://blackhawk.hepforge.org/
\end{abstract}

\maketitle

\tableofcontents
\newpage

\section{Introduction}
Black Holes (BHs) provide a unique framework for testing our understanding of the Universe \cite{Cardoso:2019rvt}. The vast array of observational signatures collected over the past decades \cite{Bambi:2017iyh}, together with recent experimental breakthroughs—such as the direct observation of supermassive BH shadows by the Event Horizon Telescope \cite{EventHorizonTelescope:2019dse,EventHorizonTelescope:2021dqv,EventHorizonTelescope:2022xqj} and the detection of gravitational waves from compact binary mergers by the LIGO/Virgo/KAGRA collaborations \cite{LIGOScientific:2016aoc,LIGOScientific:2017vwq,LIGOScientific:2018dkp}, have transitioned BHs from hypothetical objects to primary laboratories for testing gravity and fundamental physics in the strong-field regime, where departures from General Relativity (GR) might first manifest~\cite{Creminelli:2017sry,Sakstein:2017xjx,Ezquiaga:2017ekz,Boran:2017rdn,Baker:2017hug,Amendola:2017orw,Visinelli:2017bny,Crisostomi:2017lbg,Dima:2017pwp,Cai:2018rzd,Casalino:2018tcd,Barack:2018yly,LIGOScientific:2018dkp,Casalino:2018wnc,Held:2019xde,Bambi:2019tjh,Vagnozzi:2019apd,Zhu:2019ura,Cunha:2019ikd,Banerjee:2019nnj,Banerjee:2019xds,Allahyari:2019jqz,Vagnozzi:2020quf,Khodadi:2020jij,Kumar:2020yem,Khodadi:2020gns,Pantig:2021zqe,Khodadi:2021gbc,Roy:2021uye,Uniyal:2022vdu,Pantig:2022ely,Ghosh:2022kit,Khodadi:2022pqh,KumarWalia:2022aop,Shaikh:2022ivr,Odintsov:2022umu,Oikonomou:2022tjm,Pantig:2023yer,Gonzalez:2023rsd,Sahoo:2023czj,Nozari:2023flq,Uniyal:2023ahv,Filho:2023ycx,Raza:2023vkn,Hoshimov:2023tlz,Chakhchi:2024tzo,Liu:2024lbi,Liu:2024lve,Khodadi:2024ubi,Liu:2024soc,Nojiri:2024txy}. The success of these observational milestones, alongside the development of future missions—including next-generation ground-based gravitational-wave detectors like the Einstein Telescope \cite{ET:2019dnz,ET:2025xjr} and Cosmic Explorer \cite{Reitze:2019iox,Evans:2021gyd}, the space-based LISA observatory \cite{LISACosmologyWorkingGroup:2022jok}, the Next-Generation EHT (ngEHT) \cite{Tiede:2022grp,Doeleman:2023kzg,Johnson:2023ynn}, and the Black Hole Explorer \cite{Johnson:2024ttr}, see \cite{Genzel:2024vou} for a comprehensive overview—has also renewed interest in more speculative scenarios that could become observationally accessible in the near future. 

Among these, Primordial Black Holes (PBHs) have arguably attracted the most widespread attention as uniquely versatile probes of both early-Universe cosmology and fundamental physics, see Refs.~\cite{Khlopov:2008qy,Carr:2016drx,Green:2020jor,Carr:2020xqk,Villanueva-Domingo:2021spv,Carr:2021bzv,Bird:2022wvk,Carr:2023tpt,Arbey:2024ujg,Choudhury:2024aji} for reviews. PBHs are hypothetical objects formed in the very early Universe through the direct collapse of large density perturbations upon horizon re-entry~\cite{Chapline:1975ojl,Meszaros:1975ef,Khlopov:1980mg,Khlopov:1985fch,Ivanov:1994pa,Choudhury:2013woa,Belotsky:2014kca,Bird:2016dcv,Clesse:2016vqa,Poulin:2017bwe,Raccanelli:2017xee,LuisBernal:2017fmf,Clesse:2017bsw,Kohri:2018qtx,Liu:2018ess,Liu:2019rnx,Murgia:2019duy,Carr:2019kxo,Liu:2020cds,Hertzberg:2020hsz,Serpico:2020ehh,DeLuca:2020bjf,DeLuca:2020fpg,DeLuca:2020qqa,Carr:2020erq,Bhagwat:2020bzh,DeLuca:2020sae,Wong:2020yig,Carr:2020mqm,Domenech:2020ssp,DeLuca:2021wjr,Arbey:2021ysg,Franciolini:2021tla,DeLuca:2021hde,Cheek:2021odj,Cheek:2021cfe,Heydari:2021gea,Dvali:2021byy,Heydari:2021qsr,DeLuca:2021pls,Liu:2021jnw,Saha:2021pqf,Bhaumik:2022pil,Anchordoqui:2022txe,Cai:2022erk,Oguri:2022fir,Franciolini:2022tfm,Mazde:2022sdx,Cai:2022kbp,Anchordoqui:2022tgp,Liu:2022iuf,Fu:2022ypp,Choudhury:2023vuj,Papanikolaou:2023crz,deFreitasPacheco:2023hpb,Choudhury:2023jlt,Choudhury:2023rks,Musco:2023dak,Yuan:2023bvh,Choudhury:2023hvf,Ghoshal:2023sfa,Cai:2023uhc,Choudhury:2023kdb,Huang:2023chx,Choudhury:2023hfm,Bhattacharya:2023ysp,Heydari:2023xts,Heydari:2023rmq,Choudhury:2023fwk,Choudhury:2023fjs,Ghoshal:2023pcx,Khan:2025kag,Mittal:2021egv,Hai-LongHuang:2023atg,Huang:2023mwy,Anchordoqui:2024akj,Choudhury:2024one,Thoss:2024hsr,Papanikolaou:2024kjb,Choudhury:2024ybk,Choudhury:2024jlz,Anchordoqui:2024dxu,Papanikolaou:2024fzf,Yin:2024xov,Choudhury:2024dei,Heydari:2024bxj,Dvali:2024hsb,Boccia:2024nly,Huang:2024aog,Choudhury:2024dzw,Anchordoqui:2024jkn,Yang:2024vij,Saha:2024ies,Anchordoqui:2024tdj,Chen:2024pge,Dai:2024guo,Huang:2024koy,Zantedeschi:2024ram,Chianese:2024rsn,Barker:2024mpz,Borah:2024bcr,Hai-LongHuang:2024gtx,Calza:2021czr,Calza:2022ljw,Calza:2023rjt,Calza:2023gws,Calza:2023iqa,Calza:2025mrt,Calza:2025yfm,Calza:2022ioe}. Although currently undetected, they remain of significant theoretical interest. Indeed, forming deep in the radiation-dominated era, PBHs evade Big Bang Nucleosynthesis (BBN) constraints on the baryon abundance, making them viable cold Dark Matter (DM) candidates, particularly within the so-called asteroid-mass window, $10^{17}\,{\text{g}} \lesssim M_{\text{pbh}} \lesssim 10^{23}\,{\text{g}}$~\cite{Katz:2018zrn,Bai:2018bej,Smyth:2019whb,Coogan:2020tuf,Ray:2021mxu,Auffinger:2022dic,Ghosh:2022okj,Miller:2021knj,Branco:2023frw,Bertrand:2023zkl,Tran:2023jci,Gorton:2024cdm,Dent:2024yje,Tamta:2024pow,Tinyakov:2024mcy,Loeb:2024tcc}. This possibility has been widely discussed in the literature, yielding numerous constraints on PBH abundance across different mass scales. Indeed, PBHs can span a vast mass range, determined solely by their formation time, and, quite interestingly, those formed at the earliest epochs would be light enough to exhibit quantum properties. Incorporating these quantum effects in highly curved spacetimes led Hawking to discover black hole evaporation, establishing a profound connection between gravity, thermodynamics, and quantum mechanics \cite{Hawking:1974rv, Hawking:1974sw, Hawking:1975iha,Hawking:1975vcx}. In the semiclassical regime, particle emission causes a BH to lose mass and eventually evaporate completely. Because the Hawking temperature is inversely proportional to the BH mass, the temperature rises during evaporation, leading to the emission of increasingly energetic particles and, theoretically, the entire spectrum of fundamental degrees of freedom \cite{Page:1976df,Page:1976ki,Page:1977um}. This continuous process yields potentially observable signatures in gamma rays, cosmic rays, neutrinos, or cosmological observables like BBN and the Cosmic Microwave Background (CMB), offering novel avenues to search for dark sectors and physics Beyond the Standard Model (BSM) \cite{Carr:2020gox,Calza:2021czr,Calza:2022ljw,Calza:2023rjt,Calza:2023gws,Calza:2023iqa,Perez-Gonzalez:2020vnz,Bernal:2020bjf,Bernal:2020ili,Bernal:2021bbv,Bernal:2021yyb,Bernal:2022oha,Perez-Gonzalez:2023uoi,Cheek:2021cfe,Cheek:2021odj,Cheek:2022dbx,Cheek:2022mmy,Bertuzzo:2024fns,DeRomeri:2024zqs,Gunn:2024xaq,Perez-Gonzalez:2025try,Sanchis:2025awq,Airoldi:2025bgr,Airoldi:2025opo,IguazJuan:2025vmd}.

These phenomenological prospects necessitate robust numerical tools capable of accurately tracking complex decay cascades and computing the final particle emission spectra originating from Hawking evaporation. To this end, the public code \texttt{BlackHawk} was introduced as a flexible framework for calculating these spectra for generic BH populations, enabling users to constrain the PBH dark matter fraction or predict Hawking radiation signals for future instruments \cite{Arbey:2019mbc,Arbey:2021mbl}. Historically, Hawking radiation was also considered a means to detect microscopic BHs evaporating in high-energy colliders like the LHC, prompting the development of earlier codes such as \texttt{BlackMax} \cite{Dai:2009by} and \texttt{Charybdis} \cite{Frost:2009cf}. Previous versions of \texttt{BlackHawk} implemented the evaporation of Schwarzschild, Kerr, charged, higher-dimensional, and polymerized BHs, alongside the computation of primary and secondary spectra for both Standard Model and BSM particles. 

Recently, some of the authors proved that considering \textit{regular} BH (RBH) metrics can significantly alter PBH evaporation constraints, indicating that the theoretical resolution of spacetime singularities has direct, calculable phenomenological consequences for the interplay between BHs and DM \cite{Calza:2022ioe,Calza:2024fzo,Calza:2024xdh,Calza:2025mwn,Calza:2025yfm}. Born from the expectation that quantum gravity must smooth out the singularities predicted by classical theory, these geometries typically substitute the central singularity with a regular core, such as a localized de Sitter vacuum or a wormhole throat. Consequently, they serve as excellent phenomenological prototypes to explore how microscopic quantum corrections manifest macroscopically in black hole spacetimes. The findings of~\cite{Calza:2022ioe,Calza:2024fzo,Calza:2024xdh,Calza:2025mwn,Calza:2025yfm} highlighted the need for \texttt{BlackHawk v3.0}, a comprehensive tool designed to describe the Hawking radiation of \textit{non-singular} and \textit{beyond-Schwarzschild} geometries. Building on previous iterations, v3.0 introduces several singularity-free metrics, focusing on regular and quantum-gravity-inspired spacetimes. Specifically, it incorporates the Bardeen \cite{Bardeen:1968ghw} and Hayward \cite{Hayward:2005gi} regular BHs, the Simpson--Visser \cite{Simpson:2018tsi} and Peltola--Kunstatter \cite{Peltola:2008pa} black bounces, the D'Ambrosio--Rovelli \cite{DAmbrosio:2018wgv} spacetime, and the Babichev--Charmousis--Leh'ebel (BCL) \cite{Babichev:2017guv} BH, an exact solution within a subclass of Horndeski theories of particular interest for the study of quasinormal modes \cite{Arbey:2025ses}. For these models, the code computes Hawking spectra using dedicated numerical routines from the companion code \texttt{GrayHawk} \cite{Calza:2025whq}, which evaluates the GreyBody Factors (GBFs) that quantify the energy-dependent departure of the emission from a pure blackbody spectrum \cite{Calza:2025whq}. It is worth mentioning that {\tt GrayHawk v2} \cite{Calza:2026wuf}, which enables the user to consider the scattering of waves on many black hole and wormhole metrics in a semi-analytical or fully numerical way, is twinned-released with the release of {\tt BlackHawk v3.0}. Beyond the new metrics, { \tt BlackHawk v3.0} features significant technical optimizations, as e.g. the ones described in \cite{Arbey:2025dnc}, improving interpolation routines, data table handling, and overall code efficiency, thereby facilitating future extensions to more general BH models and BSM scenarios. \texttt{BlackHawk v3.0} is publicly available at
\begin{center}
~\url{https://blackhawk.hepforge.org/}.
\end{center}
This paper is organized as follows. In Sec.~\ref{sec:new_features}, we detail the new metrics implemented in \texttt{BlackHawk v3.0}, summarizing their primary geometric and thermodynamic properties, alongside the main technical improvements introduced in the code. In Sec.~\ref{sec:parameters}, we list the new input parameters available to the user. Section~\ref{sec:validation} is dedicated to code validation, where we compare our results against established literature benchmarks to verify the accuracy of the new numerical routines. Finally, we summarize our conclusions and discuss future perspectives in Sec.~\ref{sec:conclusion}. Unless otherwise specified, we use geometrized units with $G = c = k_\text{B} =\hbar = 4\pi\varepsilon_0 = 1$.

\section{New features}
\label{sec:new_features}
This Section is devoted to the description of the new features of \texttt{BlackHawk v3.0}. We grouped them into two categories: \textit{Metrics} and \textit{Code improvements}. Some of these add-ons imply a modification of some parameters and routines, which are listed in Section~\ref{sec:parameters}.
\subsection{Metrics}
\label{sec:rbh}
It is well known that, in General Relativity (GR), continuous gravitational collapse of matter, under physically reasonable energy conditions, leads to spacetime \textit{singularities} ~\cite{Penrose:1964wq,Hawking:1970zqf}. Curvature singularities, characterized by the divergence of curvature invariants and geodesic incompleteness, lead to the breakdown of predictability and, presumably, of the classical theory itself. Generally, singularities are regarded as indicators that GR is incomplete and should be replaced, at sufficiently high energies, by a quantum theory of gravity. This motivated the construction of numerous models in which the singularity is replaced by a regular portion of spacetime. Those objects are dubbed Regular BHs (RBHs), if the central singularity is replace by a smooth core, most commonly de-Sitter, or Black Bounces (BBs), if the singularity is avoided by altering the internal topology, by adding a ``bounce" or a throat that connects to another region of spacetime \cite{Borde:1996df,AyonBeato:1998ub,AyonBeato:1999rg,Bronnikov:2005gm,Berej:2006cc,Bronnikov:2012ch,Rinaldi:2012vy,Stuchlik:2014qja,Schee:2015nua,Johannsen:2015pca,Myrzakulov:2015kda,Fan:2016hvf,Sebastiani:2016ras,Toshmatov:2017zpr,Chinaglia:2017uqd,Frolov:2017dwy,Bertipagani:2020awe,Nashed:2021pah,Simpson:2021dyo,Franzin:2022iai,Chataignier:2022yic,Ghosh:2022gka,Khodadi:2022dyi,Pedrotti:2025idg, Sharif:2022yxg,Wahlang:2017zvk,Farrah:2023opk,Fontana:2023zqz,Boshkayev:2023rhr,Luongo:2023jyz,Luongo:2023aib,Cadoni:2023lum,Giambo:2023zmy,Cadoni:2023lqe,Luongo:2023xaw,Sajadi:2023ybm,Javed:2024wbc,Ditta:2024jrv,Al-Badawi:2024lvc,Ovgun:2024zmt,Corona:2024gth,Bueno:2024dgm,Konoplya:2024hfg,Pedrotti:2024znu,Bronnikov:2024izh,Kurmanov:2024hpn,Bolokhov:2024sdy,Agrawal:2024wwt,Belfiglio:2024wel,Stashko:2024wuq,Faraoni:2024ghi,Konoplya:2024lch,Khodadi:2024efq,Calza:2024qxn, Calza:2024fzo, Calza:2024xdh,Yuan:2025eyi,Calza:2025mwn}. 

RBHs and BBs attracted increasing attention in recent years after the development of several frameworks like non-linear electrodynamics, effective quantum gravity corrections, polymer quantization, modified gravity theories, or phenomenological regularization prescriptions. As shown in \cite{Calza:2024fzo,Calza:2024xdh,Calza:2025mwn}, such spacetime regularizations incur modifications in both the Hawking temperature and the greybody factors, thus changing the phenomenology relative to the Schwarzschild case. Many RBHs and BBs are characterized by a decrease in the Hawking temperature with respect to the standard case as the regularizing parameter increases. This leads to slower evaporation and potentially longer-lived PBHs. At the same time, non-trivial modifications of the emitted spectra may be generated by the different greybody factors' shapes that may enhance or dim specific energy ranges. Consequently, constraints on PBH abundances derived from evaporation observables can be substantially modified when regular metrics are considered \cite{Calza:2024fzo,Calza:2024xdh,Calza:2025mwn}. These results suggest that the resolution of spacetime singularities may have direct phenomenological implications for the interplay between BH physics, dark matter, and quantum gravity. 

Since the publication of its last versions \cite{Arbey:2019mbc, Arbey:2021mbl}, we updated \texttt{BlackHawk} by adding some new singularity-free metrics: the BCL BH metric \cite{Babichev:2017guv}, an exact solution of a special subclass of Horndeski theories, two regular BH metrics, namely the Bardeen and Hayward BHs, two black bounces, the Simpson-Visser and Peltola-Kunstatter BB, and a black hole-to-white hole geometry, i.e., the D'Ambrosio-Rovelli metric. 

All the metrics considered here enjoy spherical symmetry, and their line element can be written in the following general form, in Boyer-Lindquist coordinates:
\begin{equation}
ds^2 = -G(r)dt^2+F(r)^{-1}dr^2+H(r)d\Omega^2,
\label{eq:intrinsic}
\end{equation}
where $d\Omega^2 = d\theta^2 + \sin^2\theta d\varphi^2$ is the metric on the 2-sphere and the metric functions $F(r),\,G(r),$ and $H(r)$ are such that
\begin{equation}
G(r) \xrightarrow{r \to \infty} 1\,, \quad F(r) \xrightarrow{r \to \infty} 1\,, \quad H(r) \xrightarrow{r \to \infty} r^2\,,
\label{eq:asflat}
\end{equation}
namely, the spacetimes are asymptotically flat. BHs whose line element can be recast in the form of Eq.~(\ref{eq:intrinsic}) in Boyer-Lindquist coordinates fall into the Petrov-D type category \cite{Kinnersley:1969zza}; we refer the interested reader to \cite{Petrov:2000bs,Coley:2004jv} for a comprehensive discussion on the Petrov classification.

We now introduce an important quantity for what follows, namely the Hawking temperature $T$ of a black hole, which is defined as 
\begin{equation}
    T=\sqrt{\frac{F(r)}{G(r)}}\frac{G'(r)}{4\pi}\Bigg|_{r_\text{H}}\,,
    \label{eq:T}
\end{equation}
where we adopted the Gibbons–Hawking definition, thereby implicitly assuming a standard Boltzmann–Gibbs distribution. In Eq.~(\ref{eq:T}) $'$ denotes differentiation with respect to $r$, while $r_\text{H}$ is the event horizon radius, which is defined as the largest real root of $F(r)=0$. In what follows, we present a list of the new BH metrics implemented in \texttt{BlackHawk v3.0}. We note that the functions $F(r)$, $G(r)$, and $H(r)$ depend on the radial coordinate and on a regularizing parameter $\ell$, ranging from a minimum value $\ell=0$, which recovers the Schwarzschild solution, and a maximum value $\ell_ {Max}$ which in general differs for different BHs. For this reason, we normalized the regularizing parameter by defining
\begin{equation}
    \mathfrak{l}\equiv\ell/\ell_{\text{max}} \in [0,1).
\end{equation}
Because the physical interpretation of regularizing parameters differs across models, the treatment of BH evolution in regular spacetimes requires particular care. In some constructions, the regularizing parameter may correspond to a conserved charge or a dynamical quantity that evolves during evaporation; in others, it emerges as an effective parameter inherited from an underlying high-energy theory and may remain approximately constant throughout the evaporation process. Given the present theoretical uncertainties, \texttt{BlackHawk v3.0} adopts a conservative and agnostic approach: while instantaneous Hawking spectra are computed for all implemented metrics, the code, in its current version, does not attempt to model the time evolution of BHs unless a consistent dynamical framework is available.

\subsubsection{Bardeen Black Hole}
The Bardeen BH~\cite{Bardeen:1968ghw} is one of the most well-known examples of RBHs and, to the best of our knowledge, the first RBH that appeared in the literature. It is defined by the line element
\begin{equation}
    ds^2 = -\left(1 - \frac{2Mr^2}{(r^2+\ell^2)^{3/2}}\right)dt^2+\left(1 - \frac{2Mr^2}{(r^2+\ell^2)^{3/2}}\right)^{-1}dr^2+r^2d\Omega^2\,,
    \label{eq:bardeen}
\end{equation}
where $\ell$ is the regularizing parameter and $M$ denotes the BH mass.\footnote{For all metrics considered in this work, the parameter $M$ appearing in the metric function can be unambiguously identified with the BH mass (Komar, ADM, Misner–Sharp–Hernandez, or Brown–York). } 
It is easy to see that the Schwarzschild limit is smoothly recovered as $\ell \to 0$. For the metric Eq.~(\ref{eq:bardeen}) to describe a BH rather than a horizonless object, the regularizing parameter $\ell$ must satisfy $\ell \leq \sqrt{16/27}\,M \approx 0.77\,M$. An alternative, arguably more physically motivated, parametrization is obtained by expressing the regularizing parameter as a fraction of the horizon radius $r_H$. In this reparametrization, $\ell$ is constrained to be $\ell \lesssim 0.70\,r_H$, a limit which is saturated in the regime where the temperature vanishes. 
%
A notable feature of the Bardeen BH is the presence of a de Sitter (dS) core that replaces the Schwarzschild singularity. Indeed, as $r \to 0$, the metric function behaves as $G(r)=F(r) \propto r^2$, consistent with the structure of an asymptotically dS spacetime. Although initially introduced as a phenomenological model, the Bardeen BH has since been shown to arise as a magnetic monopole configuration~\cite{Ayon-Beato:2000mjt} within specific nonlinear electrodynamics models~\cite{Ayon-Beato:2004ywd}. Alternative origins have also been proposed, including quantum corrections to the uncertainty principle~\cite{Maluf:2018ksj}. \\
The temperature of the Bardeen BH, given as a function of $\mathfrak{l} \equiv \sqrt{27/16} \, \ell/M$, so that $\mathfrak{l} \in \left[0,1\right)$, is given by:
\begin{equation}
    T = \frac{1}{4\pi r_H}\frac{r_H^2 - \frac{32}{27}M^2\mathfrak{l}^2}{\left(r_H^2 + \frac{16}{27}M^2\mathfrak{l}^2 \right)},
    \label{eq:T_bardeen}
\end{equation}
where $r_H$ is obtained by solving
\begin{equation}
    \left(r_H^2 + \frac{16}{27}M^2\mathfrak{l}^2\right)^{3/2} = 2Mr_H^2.
\end{equation}

\subsubsection{Hayward Black Hole}
Another well-known example of RBH is the Hayward BH~\cite{Hayward:2005gi}, which is defined by the line element:
\begin{eqnarray}
    ds^2 = -\left(1 - \frac{2Mr^2}{r^3 + 2M\ell^2}\right)dt^2+\left(1 - \frac{2Mr^2}{r^3 + 2M\ell^2}\right)^{-1}dr^2+r^2d\Omega^2\,.
\end{eqnarray}
Also, in this case, it is straightforward to see that the Schwarzschild solution is smoothly recovered in the $\ell \to 0$ limit. When expressed in terms of the BH mass $M$, the regularizing parameter $\ell$ is subject to the same bound as the Bardeen BH, namely $\ell \leq \sqrt{16/27}\,M$. Alternatively, if one adopts the re-parametrization in terms of the horizon radius $r_H$, the constraint becomes $\ell \lesssim 0.57\,r_H$, which, again, corresponds to the regime where the temperature vanishes. 
Exactly as in the Bardeen BH case, the Hayward space-time features a de Sitter (dS) core that replaces the central singularity. In fact, Hayward’s original motivation was precisely to resolve the singularity by introducing a dS core with an effective cosmological constant $\Lambda = 3/\ell^2$. Despite being proposed phenomenologically, possible theoretical origins of the Hayward black hole have since been explored. These include modifications to the equation of state of matter at high densities~\cite{Sakharov:1966aja,1966JETP...22..378G}, finite-density and finite-curvature effects~\cite{1982JETPL..36..265M,1987JETPL..46..431M,Mukhanov:1991zn}, models based on nonlinear electrodynamics~\cite{Kumar:2020bqf,Kruglov:2021yya}, and, more broadly, quantum-gravity–induced corrections~\cite{Addazi:2021xuf,AlvesBatista:2023wqm}.\\
The temperature of the Hayward BH, given as a function of $\mathfrak{l} \equiv \sqrt{27/16} \, \ell/M$, so that $\mathfrak{l} \in \left[0,1\right)$, is given by:
\begin{equation}
    T = \frac{1}{4\pi r_H}\left(1 - \frac{16}{9}\frac{\mathfrak{l}^2M^2}{r_H^2}\right)
    \label{eq:T_hayward}
\end{equation}
where $r_H$ is obtained by solving
\begin{equation}
    r_H^3 = 2 M r_H^2 - \frac{32}{27} M^3\mathfrak{l}^2.
\end{equation}
\subsubsection{Simpson-Visser Black Bounce}
The Simpson–Visser metric is a one-parameter extension of the Schwarzschild metric and is arguably the most popular example of black-bounce geometry. As Simpson and Visser themselves describe it, this construction “represents the minimal violence to the standard Schwarzschild solution" required to enforce regularity~\cite{Simpson:2018tsi}. The line element of the Simpson-Visser spacetime reads 
\begin{equation}
    ds^2 = -\left(1 - \frac{2M}{\sqrt{r^2 + \ell^2}} \right) dt^2 
+ \frac{dr^2}{1 - \frac{2M}{\sqrt{r^2 + \ell^2}}} 
+ (r^2 + \ell^2)\, d\Omega^2 .
\end{equation}
The Simpson-Visser space-time exhibits a rich phenomenology, smoothly interpolating between a BH and a traversable wormhole, depending on the value of the regularizing parameter; in particular
\begin{itemize}
    \item for $\ell = 0$, it reduces to the Schwarzschild BH;
    \item for $0 < \ell < 2 M$, it describes a RBH with a one-way spacelike throat;
    \item for $\ell = 2M$, it yields a one-way wormhole with an extremal null throat;
    \item  for $\ell > 2M$, it describes a traversable wormhole with a two-way timelike throat.
\end{itemize}
Since its introduction, the Simpson-Visser metric has been the subject of numerous follow-up studies (see, for example Refs.~\cite{Tsukamoto:2020bjm,Mazza:2021rgq,Shaikh:2021yux,Islam:2021ful,Guerrero:2021ues,Bambhaniya:2021ugr,Yang:2022xxh,Riaz:2022rlx,Arora:2023ltv,Jha:2023wzo,Jha:2023nkh}). Although originally motivated from a phenomenological perspective, it can also arise as a solution of general relativity coupled to nonlinear electrodynamics and a minimally coupled phantom scalar field~\cite{Bronnikov:2021uta}.\\
The temperature of the Simpson-Visser BH, given as a function of $\mathfrak{l} \equiv \frac{\ell}{2M}$, is given by:
\begin{equation}
    T = \frac{\sqrt{1-\mathfrak{l}^2}}{8\pi M}
    \label{eq:T_SV}
\end{equation}
using the fact that
\begin{equation}
    r_H = 2M \sqrt{1 -\mathfrak{l}^2}.
\end{equation}
\subsubsection{Peltola-Kunstatter Black Bounce}
The Peltola–Kunstatter space-time is a loop quantum gravity (LQG)–inspired metric obtained by applying effective polymerization techniques to the Schwarzschild BH. While LQG is expected to resolve the singularities inherent in general relativity, the complexity of the full quantum system motivates the use of semiclassical polymer quantization methods. These provide an alternative to standard Schrödinger quantization—unitarily inequivalent but retaining the key feature of spacetime discreteness. In the Peltola–Kunstatter construction, polymerization is applied to the area variable but not to the conformal mode. This leads to a geometry in which the central singularity is replaced by a complete, nonsingular bounce: the spacetime contracts to a minimum radius before re-expanding into a Kantowski–Sachs universe~\cite{Peltola:2008pa,Peltola:2009jm}. The line element reads
\begin{equation}
ds^2 = -\left(\frac{r-2M}{\sqrt{r^2+\ell^2}}\right) dt^2 
+ \frac{dr^2}{\frac{r-2M}{\sqrt{r^2+\ell^2}}} 
+ (r^2+\ell^2) d\Omega^2 .
\end{equation}
The Peltola-Kunstatter space-time is an example of a RBH metric derived from quantum gravity considerations, in contrast to the more phenomenologically motivated constructions discussed earlier.\\
The temperature of the Peltola-Kunstatter BH does not have a maximum value for its regularizing parameter. To maintain consistency with the other choices, we took a value of $\ell_{Max}=2M$ and, therefore, expressed $\mathfrak{l} \equiv \ell/2M$, so that $\mathfrak{l} \in \left[0,1\right)$, and the temperatures read:
\begin{equation}
    T = \frac{1}{8\pi M\sqrt{1 + \mathfrak{l}^2}}
    \label{eq:T_PK}
\end{equation}
using the fact that $r_H = 2M$.
\subsubsection{D'Ambrosio-Rovelli Black Hole-to-White Hole Metric}
The D’Ambrosio–Rovelli space-time, also motivated by LQG, was originally developed with aims other than singularity resolution. It provides a natural extension of the Schwarzschild geometry in which the $r=0$ singularity is smoothly crossed into the interior of a white hole, and can be regarded as the $\hbar \to 0$ limit of an effective quantum gravity metric~\cite{Bianchi:2018mml,DAmbrosio:2018wgv}. This black hole–to–white hole tunnelling mechanism has been suggested as a possible avenue toward resolving the information loss paradox. For our purposes, the key property is that the D'Ambrosio-Rovelli metric is regular. The line element reads:
\begin{equation}
ds^2 = -\left(1 - \frac{2M}{\sqrt{r^2+\ell^2}}\right) dt^2 
+ \frac{dr^2}{1 - \frac{2M}{\sqrt{r^2+\ell^2}}} \left(1 + \frac{\ell}{\sqrt{r^2+\ell^2}}\right) 
+ (r^2+\ell^2) d\Omega^2 .
\end{equation}
As with the Peltola-Kunstatter space-time, the D'Ambrosio-Rovelli metric is a well-motivated, quantum gravity–inspired example of a regular geometry. Notably, the existence of primordial D'Ambrosio-Rovelli BH would necessarily imply the presence of long-lived primordial white holes, arising from quantum transitions near the would-be singularity. This could give rise to an intriguing phenomenology, though a detailed investigation lies beyond the scope of this work.\\
The temperature of the D'Ambrosio-Rovelli BH, given as a function of $\mathfrak{l} \equiv \frac{\ell}{2M}$, so that $\mathfrak{l} \in \left[0,1\right)$, is given by:
\begin{equation}
    T = \frac{\sqrt{1 - \mathfrak{l}}}{8\pi M}
    \label{eq:T_DAR}
\end{equation}
using the fact that
\begin{equation}
    r_H = 2M \sqrt{1 -\mathfrak{l}^2}.
\end{equation}
\subsubsection{Babichev-Charmousis-Lehébel Black Hole}

The Babichev-Charmousis-Lehébel (BCL) metric \cite{Langlois:2021xzq,Babichev:2017guv} is a black hole solution of a subset of Horndeski theories referred to as the quadratic shift-symmetric theories, which are described by the action
\begin{equation}
    S[g_{\mu\nu},\phi] = \int d^4x \left[F(X)R + P(X) + Q(X)\square\phi + 2 \frac{\partial F}{\partial X}\left(\phi_{\mu\nu}\phi^{\mu\nu} - (\square\phi)^2\right)\right],
    \label{eq:action}
\end{equation}
where $X \equiv \nabla^{\mu}\phi\nabla_{\mu}\phi$ and $\phi_{\mu\nu}\equiv \nabla_{\mu}\nabla_{\nu}\phi$. The action Eq.~(\ref{eq:action}) features no cubic terms in the second derivative of the scalar field $\phi$, and it is invariant under constant shifts of the scalar field $\phi$ itself, hence the name. The BCL black hole is a solution of the theory in the special case when
\begin{equation}
    F(X) = F_0 + F_1\sqrt{X}, \qquad P(X) = -P_1X, \qquad Q(X)=0,
    \label{eq:theory_func}
\end{equation}
with $F_0,F_1,P_0$ being constants. The BCL solution is described by the line element (\ref{eq:intrinsic}), with the metric functions given by
\begin{equation}
G(r)= F(r) = 
     -\left(1 -\frac{r_+}{r}\right)\left(1 + \frac{r_-}{r}\right), \qquad H(r) = r^2
\end{equation}
where $r_+$ and $r_-$ are related to the functions (\ref{eq:theory_func}) by
\begin{equation}
    r_+r_-= \frac{F_1^2}{2F_0P_1}, \qquad r_+ - r_- = 2M, \qquad r_+ > r_- > 0.
\end{equation}
Interestingly, this metric can be recast in a form that resembles that of the Reissner-Nordstr\"om black hole by introducing the parameter $\ell \equiv 2\frac{r_+r_-}{2r^2} =\frac{F_1^2}{F_0P_1r_s^2}$. With this definition, the metric functions can be written as: 
\begin{equation}
    F(r) = G(r) = 1 - \frac{2M}{r} - \frac{4M^2\ell}{2r^2},
    \label{eq:RN_resemblance}
\end{equation}
We notice that, despite the resemblance made explicit in Eq. (\ref{eq:RN_resemblance}), the BCL black hole cannot be mapped into the Reissner-Nordstr\"om solution of general relativity as it features only one horizon at $r_+$. This makes the BCL BH fundamentally different from the Reissner-Nordstr\"om one which, on the contrary, is characterized by two physical horizons. We also want to stress that the BCL metric, from which the Schwarzchild metric is recovered in the limit $\xi \to 0$, is not regular, as the $r=0$ essential singularity is not cured.  Nevertheless, this BH solution stands out for its simplicity and for being one of the few solutions from modified gravity for which the quasinormal modes \cite{Kokkotas:1999bd, Berti:2009kk, Konoplya:2011qq} for both physical and test field perturbations have been extensively studied \cite{Langlois:2021xzq, Roussille:2023sdr, Arbey:2025ses}. The temperature of the BCL BH, given as a function of $\mathfrak{l} \equiv \frac{r_-}{r_+}$, 
so that $\mathfrak{l} \in \left[0,1\right]$, is given by:
\begin{equation}
    T = \frac{\mathfrak{l}^2-1}{8\pi M}
    \label{eq:T_BCL}
\end{equation}
where we used the fact that
\begin{equation}
    r_H = \frac{2M}{1-\mathfrak{l}}.
\end{equation}

\begin{figure}[h!]
    \centering
    \includegraphics[width=0.5\linewidth]{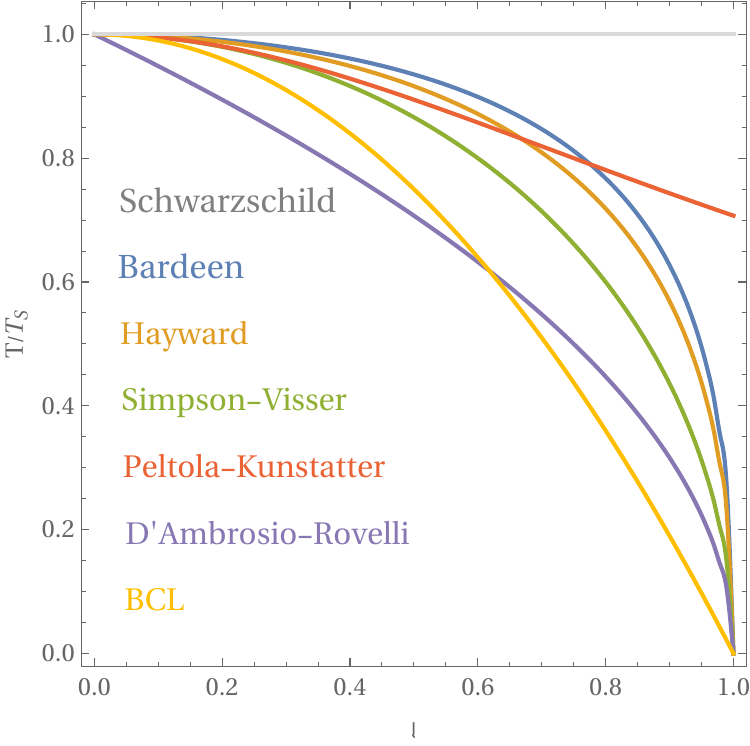}
    \caption{Temperature dependence on the $\mathfrak{l}$ parameters for the different types of BHs included in {\tt BlackHawk v3.0} normalized to the temperature of a Schwarzschild BH.}
    \label{fig:temperatures}
\end{figure}

%
\subsection{Code Improvements}
We discuss here the technical improvements made to the previous versions of the code \cite{Arbey:2019mbc,Arbey:2021mbl}. More specifically, we adopted a new way of computing GBFs, by means of \texttt{GrayHawk} \cite{Calza:2025whq}, a public code developed by one of us; we improved the precision of Kerr tables, following the discussion of some of us in \cite{Arbey:2025dnc}; and we improved on the interpolation routines of the previous versions of \texttt{BlackHawk}. In particular, for the interpolation of high- and low-energy extrapolations, we now interpolate on the values of the extrapolating functions instead of on their coefficients, and we checked that this leads to more stable results.

\subsubsection{Greybody Factors}

The Hawking radiation rate of a given particle species $i$ with spin $s$, as predicted by Hawking \cite{Hawking:1974rv, Hawking:1974sw, Hawking:1975iha, Hawking:1975vcx}, is given by
\begin{equation}
\frac{d^2N_i}{dtdE_i}=\frac{1}{2\pi}\sum_{l,m}\frac{n_i\Gamma^s_{l,m}(\omega)}{ e^{\omega/T}\pm 1}\,,
\label{eq:d2ndtdei}
\end{equation}
where $n_i$ is the number of degrees of freedom of the particle in question, $E_i=\omega$ is the mode frequency (in natural units), and the plus (minus) sign in the denominator is associated with fermions (bosons). $\Gamma^s_{l,m}$ are the GBFs. The latter are functions of energy/frequency and angular momentum which govern the deviation of the emitted spectrum from that of a blackbody~\cite{Sakalli:2022xrb,Konoplya:2024lir,Konoplya:2024vuj}. Although the emitted Hawking radiation at the horizon takes the blackbody form, the potential barrier due to space-time geometry will attenuate the radiation, so that an observer at spatial infinity will measure a spectrum which differs from that of a blackbody by a frequency-dependent function $\Gamma(\omega)$. GBFs can be characterized by setting up a classical scattering problem around the BH potential barrier, with boundary conditions allowing for incoming wave packets from infinity or, equivalently, due to the symmetries of the scattering problem, originating from the horizon. As mentioned in Sec.~\ref{sec:rbh}, for BHs whose metric is in the form of Eq.~(\ref{eq:asflat}), the temperature $T$, is given by Eq.~(\ref{eq:T}), leaving the GBFs as the only undetermined quantity in Eq.~(\ref{eq:d2ndtdei}). The GBFs of the newly implemented metrics, discussed in Sec.~\ref{sec:rbh}, are computed using \texttt{GrayHawk} \cite{Calza:2025whq}, a public code for calculating the GBFs of generic spherically symmetric BHs, developed by one of us. The GBFs are mathematically defined as the transmission coefficients of the scattering problem for test fields evolving on the background geometry defined by the metric tensor of a given black hole. The Newman-Penrose formalism \cite{Newman:1961qr} provides a unified framework for expressing the field equations of different spin $s$ in a single equation, known as the Teukolsky equation \cite{Teukolsky:1972my,Teukolsky:1973ha, Teukolsky:1973ddt, Teukolsky:1974yv}. After performing a variable transformation, the radial component of the Teukolsky equation can be rewritten in a particularly compact and convenient form, commonly referred to as the Schr\"odinger-like form,
\begin{eqnarray}
    \label{eq:S_like}
    \partial_{r^*}^2\psi_s(r^*) + \left[\omega^2 - V_s(r(r^*))\right]\psi_s(r^*) = 0,
\end{eqnarray}
with $r^*$ the tortoise coordinate, defined by
\begin{equation}
    \label{eq:tortoise}
    \frac{dr^*}{dr} = \frac{1}{\sqrt{FG}},
\end{equation}
where $F$ and $G$ are the metric functions defined in Eq.~(\ref{eq:asflat}).\\
The Schr\"odinger-like form of the Teukolsky equation is particularly suitable for the computation of the scattering problem since the potential $V_s(r(r^*))$ is typically vanishing on the asymptotes $r^*\to \pm\infty$, leading to plane wave asymptotic solutions $\sim e^{\pm i\omega r^*}$. Finally, imposing purely in-going boundary conditions at the event horizon and integrating out the solution, it is possible to compute the transmission coefficient or GBFs of the problem. For practical reasons, we computed the values of a single helicity/color degree of freedom emissivity
\begin{equation}\label{Qfactors}
    Q_{slm}\equiv\frac{\Gamma^s_{l,m}}{e^{E'/T - (-1)^{2s}} },\quad\text{such that}\quad Q_s = \sum_{l,m} Q_{slm},
\end{equation}
for 50 values of $0<\mathfrak{l}<0.99$ and for a range of 200 energies $0.01 \leq x \equiv Er_S \leq 5$ and $l$ up to 10. For $x$ out of this range, we fit empirical asymptotic forms to the emissivities; we refer the reader to \cite{Arbey:2019mbc,Arbey:2021mbl}.

\subsubsection{Improved Kerr tables}
In the previous versions of \texttt{BlackHawk}, the framework discussed above was generalized to rotating BHs by means of a clever choice of radial coordinates, which allows for reducing the rotating version of the Teukolsky equation into a Schr\"odinger-like equation. Specifically, the Kerr metric in Boyer-Lindquist coordinates $(t,r,\theta,\varphi)$ reads
\begin{equation}
    ds^2 = - \left(1 - \frac{2Mr}{\Sigma}\right)dt^2 - \frac{4Mar\sin^2\theta}{\Sigma}dtd\varphi + \frac{\Sigma}{\Delta} dr^2 + \Sigma d\theta^2 + \left(r^2 + a^2 + \frac{2Ma^2r\sin^2\theta}{\Sigma}\right)\sin^2\theta d\varphi^2
\end{equation}
where $M$ is the mass of the BH, $a= J/M$ is the BH angular momentum, $\Delta = r^2 +a^2 - 2Mr$, and $\Sigma = r^2a^2\cos^2\theta$. In Kerr spacetime, the Teukolsky equation for the radial component $R_s$ of a field with spin $s$ is given by
\begin{equation}
\Delta^{-s}\partial_r(\Delta^{(s+1)}\partial_rR_s)+\left[(K^2 - 2is(r-M)K)\Delta^{-1} + 4is\omega r - _sQ^m_l\right]R_s =0,
\label{eq:radial_t}
\end{equation}
where $_sQ_l^m = _sA_l^m + a^2\omega^2 - 2a\omega m$, $K = (r^2+a^2)\omega -ma$ and $_sA_l^m(a\omega)$ are the eigenvalues of the spin-weighted spheroidal harmonics. In a series of papers, Chandrasekhar and Detweiler \cite{Chandrasekhar:1975zx, Chandrasekhar:1976zz, Chandrasekhar:1977kf} proved that Eq.~(\ref{eq:radial_t}) can be greatly simplified by adopting the following change of variables
\begin{equation}
    \frac{dr^*}{dr} = \frac{\rho^2}{\Delta},
    \label{eq:chandra_coord}
\end{equation}
with $\rho^2 \equiv r^2 + a^2 - am/\omega$, and $m$ being the azimuthal number of the specific mode. Integrating Eq.~(\ref{eq:chandra_coord}) with a vanishing integration constant yields:
\begin{equation}
    r^*(r) = r + \frac{r_Sr_+ - am/\omega}{r_+ - r_-}\log\left(\frac{r}{r_+} -1\right) - \frac{r_Sr_- - am/\omega}{r_+ - r_-}\log\left(\frac{r}{r_-} -1\right),
    \label{eq:chandra_coord2}
\end{equation}
where we have defined the Schwarzschild radius $r_S = 2M$ and $r_\pm$ as the outer (event) and inner (Cauchy) horizons of the Kerr BH, which read:
\begin{equation}
    r_\pm \equiv \frac{r_S}{2}\left(1 \pm \sqrt{1 - a_*^2}\right),
\end{equation}
where $a_* \equiv a/M$. With the new radial coordinate Eq.~(\ref{eq:chandra_coord2}), the radial part of the Teusolsky equation in Kerr spacetime Eq.~(\ref{eq:radial_t}) can be recast into a Schr\"odinger-like form, as in Eq.~(\ref{eq:S_like}). However, the price to pay for this great simplification is the fact that, for $\omega < \omega_s \equiv am/(r_Sr_+)$, the function $r^*(r)$ is not invertible, a requirement that is essential for numerically solving the Teukolsky equation in its Schr\"odinger-like form. Specifically, in the abovementioned frequency regime, the function $r^*(r)$ exhibits a minimum around $r_{min}=1.5M$, which makes it globally non-invertible, but locally invertible on two monotonic branches, see Fig. 1 of \cite{Arbey:2025dnc}. In previous versions of \texttt{BlackHawk} \cite{Arbey:2019mbc,Arbey:2021mbl}, the issue of non-invertibility of $r^*(r)$ was addressed by inverting the function on the two monotonic branches, effectively bypassing the minimum $r_{min}$ and using the solution on the first branch as a boundary condition for the integration on the second branch. In a recent paper \cite{Arbey:2025dnc}, some of us showed that this method leads to numerical inaccuracies, particularly in the superradiant regime, leading to severe overestimation of the Hawking radiation spectrum for highly-spinning black holes. In the same paper, an alternative method based on re-scaling radial coordinates and employing Frobenius-like expansions was discussed, and the greater numerical stability over the previously discussed method was assessed. This new formalism, which is briefly summarized in what follows, is implemented in the new version of \texttt{BlackHawk}, for computing the GBFs of rotating BHs.\\
By introducing a rescaling of the radial coordinate
\begin{equation}
    x \equiv \frac{r-r_+}{r_+}
    \label{coord_resc}
\end{equation}
the radial part of the Teukolsky equation, Eq.~(\ref{eq:radial_t}) can be written as
\begin{equation}
    x^2(x+\tau)^2\partial_x^2R(x) + (s+1)(2x+\tau)x(x+\tau)\partial_xR(x) + V(x)R(x) =0,
    \label{eq:rescl_radial_t}
\end{equation}
where the effective potential is given by \footnote{We notice that in \cite{Calza:2023rjt,Calza:2025whq,Arbey:2025dnc,Calza:2024ncn}, there is a typo reporting Eq.~(\ref{eq:V_potential}). Specifically, the potential term contained $4 i s \omega (x+1)$ instead of the correct $4 i s \omega r_+ (x+1)$. We underline that this typo has no consequences on the conclusions provided in those papers since all computations were performed setting $r_+=1$.}
\begin{equation}
    V(x) = k^2 - is(2x+\tau)k + (4is\omega r_+(x+1)- _sQ_l^m)x(x+\tau)
    \label{eq:V_potential}
\end{equation}
being $k = (2-\tau)(\omega - m\Omega_H)r_+ + x(x+2)\omega r_+$ and $\tau = (r_+ - r_-)/r_+$. The imposition of purely ingoing boundary conditions near the event horizon can be expressed as a Frobenius series \cite{Wilson:1928ghw, Baber_Hassé_1935, Leaver:1985ax, Leaver:1986vnb, Konoplya:2023ahd, Konoplya:2023ppx,Rosa:2012uz, Rosa:2016bli} 
\begin{equation}
    R_s(x) = x^{-s-\frac{i\varpi}{\tau}}\sum_{n=0}^{\infty}a_nx^n
    \label{eq:fro_series}
\end{equation}
which is typically used for solving second-order differential equations \cite{Calza:2024fzo,Calza:2024xdh}. Here $\varpi \equiv (2-\tau)(\bar{\omega} - m\bar{\Omega}_H)$ and barred quantities are normalized to the horizon radius, e.g., $\bar{\omega} = \omega r_+$. The asymptotic form of the solution, far from the BH, takes the form \cite{Teukolsky:1972my}: 
\begin{equation}
    R_s(x) \sim \frac{R_{s,in}^{lm}}{r_+}\frac{e^{-i\bar{\omega}x}}{x} + \frac{R_{s,out}^{lm}}{r_+^{2s+1}}\frac{e^{i\bar{\omega}x}}{x^{2s+1}}.
    \label{eq:asympt}
\end{equation}

The normalization of the scattering problem is set by choosing $a_0=1$ in Eq.~(\ref{eq:fro_series}). The GBFs can then be computed from the $R_{s,in}^{lm}(\omega)$ coefficient as:
\begin{equation}
    \Gamma^s_{lm}(\omega) = \delta_s|R_{s,in}^{lm}(\omega)|^{-2},
    \label{eq:gfb_fr}
\end{equation}
where the coefficient $\delta_s$ is given by:
\begin{equation}
    \delta_s=- i e^{i \pi s} \bar \omega^{(2 s - 1) } \left(\frac{1}{2}\right)^{1-2s} \frac{\Gamma(1-s+i2\frac{\varpi}{\tau})}{\Gamma(s+i2\frac{\varpi}{\tau})}\tau 
\end{equation}
where $\Gamma$ is Euler's gamma function. The near-horizon solution Eq.~(\ref{eq:fro_series}) is used as a boundary condition for the numerical integration of the radial Teukolsky equation up to large radial distances. Finally, to extract the value of $R_{s,in}^{lm}$, needed to compute the GBFs according to Eq.~(\ref{eq:gfb_fr}), the numerical solution at large radial distance is matched to the asymptotic form of Eq.~(\ref{eq:asympt}). This method, dubbed \textit{direct method} in \cite{Arbey:2025dnc}, was shown to perform significantly better in terms of numerical stability, as compared to the Schr\"odinger-like approach for highly spinning BHs. For this reason, in the latest version of \texttt{BlackHawk}, the \textit{direct method} has replaced the old one for the calculation of the GBFs of rotating BHs. Specifically, we used the direct method to recompute from scratch the \texttt{gamma\_tables} of Kerr BH for test fields of spin $s=0,1/2,1,3/2,2$. We finally checked the consistency of our results by comparing them against some existing benchmarks in the literature  \cite{Dong:2015yjs,Calza:2021czr,Calza:2022ljw,Calza:2023gws,Calza:2023iqa}.
\begin{figure}[t!]
    \centering
    \includegraphics[width=0.4\linewidth]{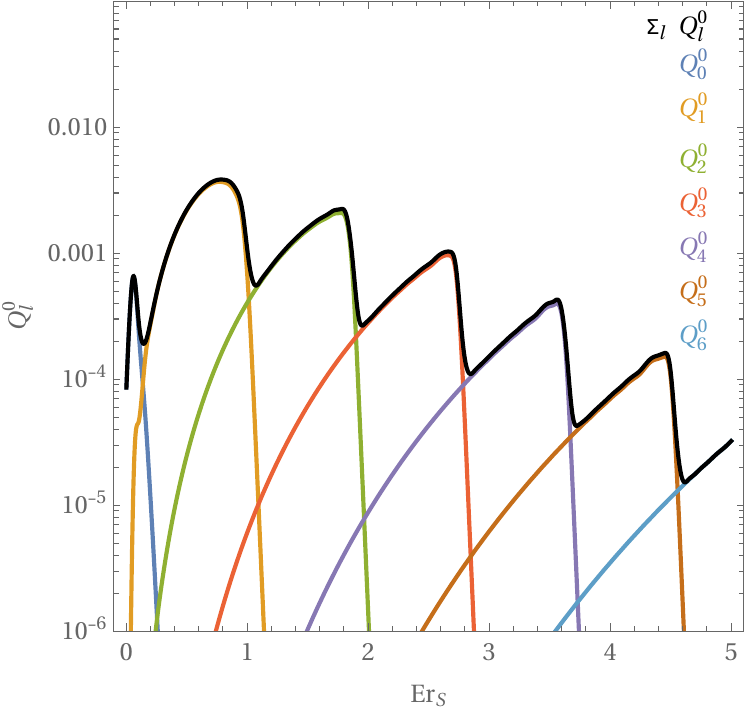}
    \includegraphics[width=0.4\linewidth]{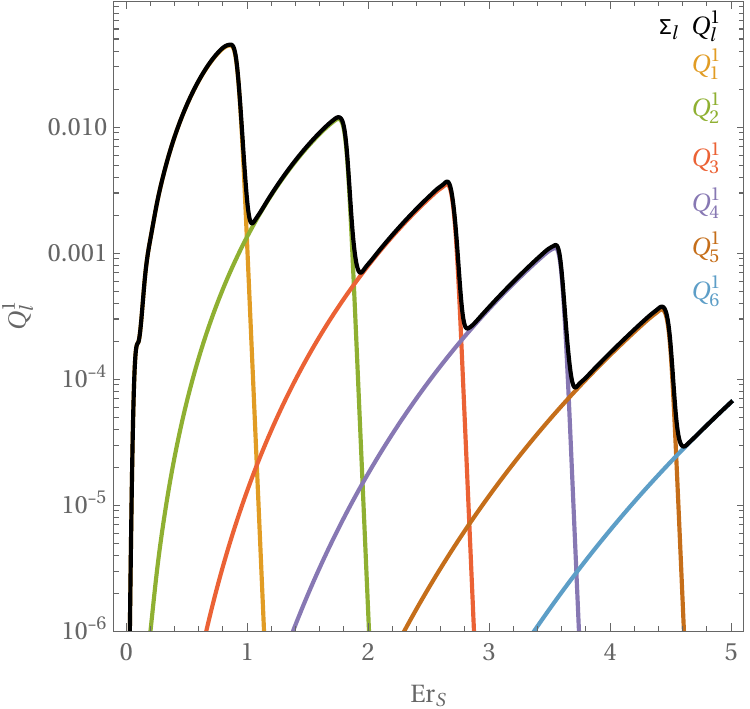}
    \caption{Emission spectra $Q^s_{l}=\sum_m Q^s_{lm}$ in the case of Kerr BHs for $s=0$ (left) and $s=1$ (right), assuming a value of $a_\star=0.99$. We present the individual contributions of the $l$ modes up to a value of $l=6$ and the \emph{total} contribution $Q_s$ obtained by summing over all modes in black.}
    \label{fig:Q_s_comparison}
\end{figure}

We present the $Q_{slm}$ for $s=0$ (left), $s=1$ (right) and $a_\star=0.99$ as function of $E r_S$ in Fig.~\eqref{fig:Q_s_comparison}.
Each $l=m$ mode is colored, while the total $Q_s$ is presented as a black line.
We observe that the contribution of the $l=m$ mode dominates the emission for highly spinning black holes, while the rest of the modes contribute in a negligible manner.
Additionally, for the scalar case, we observe that the $l=m=0$ mode contribution, not affected by the enhancement from the BH angular momentum, is about 1 order of magnitude smaller than the $l=m=1$ case, demonstrating that even for scalars, the dominating mode is the latter one.


\section{Parameters}
\label{sec:parameters}
The input parameters used by {\tt BlackHawk} are listed in a parameter file (e.g., {\tt parameters.txt} for the pre-built one). This file can be modified by the user and is saved for each new run of the code in the results/ directory.
Here we report only the newly introduced modification with respect to the previous version of {\tt BlackHawk}.
Namely, the entry {\tt metric=n} allows the selection of the newly introduced BHs. Besides the one already present in the previous version, it is now possible to choose {\tt 4=Bardeen, 5=Hayward, 6=Simpson-Visser, 7=Peltola-Kunstatter, 8=D'Ambrosio-Rovelli, 9=BCL}.
The temporal evolution of such BHs should, in principle, be characterized by the functions $M(t)$ and $\mathfrak{l}(t)$. Defining the differential equation governing $\mathfrak{l}(t)$ and relating it to $M(t)$ is model-dependent and goes beyond the scope of this code. Therefore, we decided not to include the possibility of having a temporal evolution of such newly introduced BHs. It is worth noticing that when selecting {\tt metric = 0}, the code will use the improved version of the Kerr {\tt gamma-tables}
\section{Validation}
\label{sec:validation}
In this section, we first validate {\tt BlackHawk v3.0}, comparing its results with those reported in Refs.~\cite{Calza:2024fzo,Calza:2024xdh} and, subsequently, in $\tt GrayHawk$~\cite{Calza:2025whq}. To perform such analysis, we consider the Hawking spectra of Bardeen and Hayward black holes computed in \cite{Calza:2024fzo,Calza:2024xdh,Calza:2025whq} and the corresponding ones computed with {\tt BlackHawk v3.0}, In \cite{Calza:2024fzo,Calza:2024xdh} the calculations rely on the Frobenius method to solve the scattering problem, following a reformulation of the radial equation in terms of a suitably rescaled radial coordinate. By contrast, {\tt GrayHawk} ~\cite{Calza:2025whq}, which is at the base of  {\tt BlackHawk v3.0} regular BHs,  adopts a different coordinate choice to perform the computation.

\noindent To enable a consistent comparison among these results, we first convert the values of the regularization parameter reported in Refs.~\cite{Calza:2024fzo,Calza:2024xdh}, originally expressed in units of \(r_H\), into the corresponding values in units of the black hole mass \(M\), as employed in Ref.~\cite{Calza:2025whq}. Finally, these parameters are re-expressed in units of \(\mathfrak{l}\), see Tab~\ref{tab:conversion}.

\begin{table}[ht]
\centering
\label{tab:conversion}
\begin{tabular}{l c c c}
\toprule
\textbf{Model} & \quad \; $\ell = 0.15\,r_H\,(0.293 M)$ & \qquad \; $\ell = 0.30 \,r_H\,(0.546 M)$ & \qquad \; $\ell = 0.45 \,r_H\,(0.718 M)$ \\
\midrule
Hayward & $0.381 \mathfrak{l}$ & $0.709 \,\mathfrak{l}$ & 0.$932 \,\mathfrak{l}$ \\
\addlinespace
Bardeen & $0.377\,\mathfrak{l}$ & $0.685\,\mathfrak{l}$ & $0.887 \,\mathfrak{l}$ \\
\bottomrule
\end{tabular}
\caption{Regularization parameters conversion $(\ell \rightarrow\mathfrak{l})$ for the Hayward and Bardeen RBHs.}
\end{table}

\begin{figure}[ht!]
    \centering
    \includegraphics[width=0.4\linewidth]{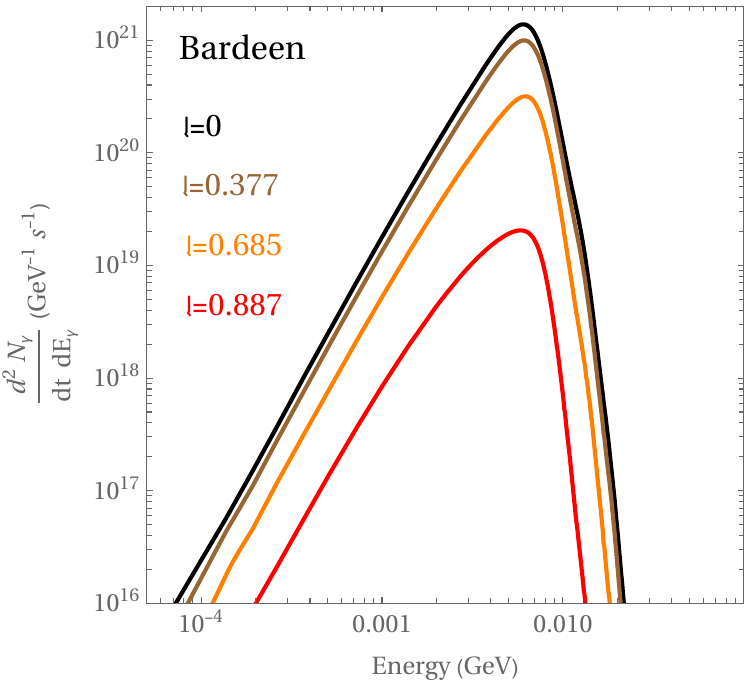}
    \includegraphics[width=0.4\linewidth]{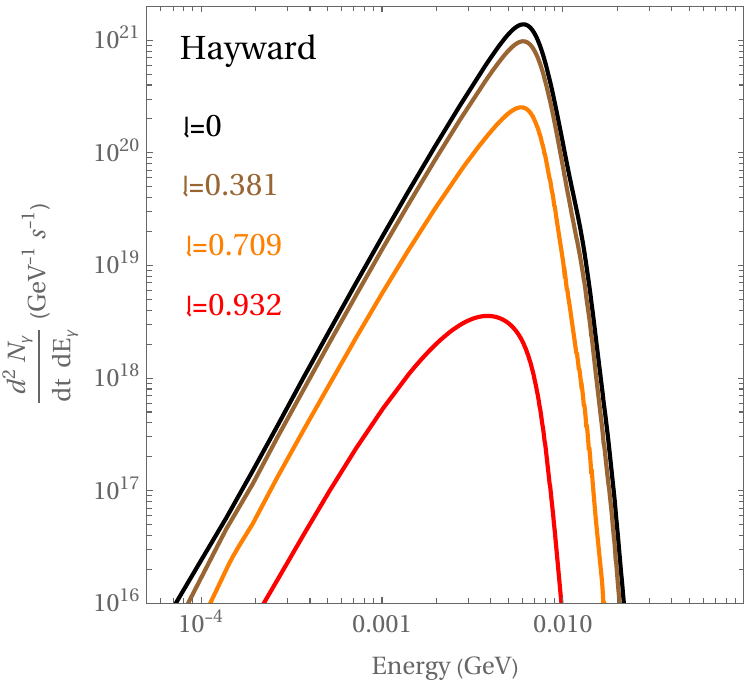}
    \caption{Emission photon spectra in the case of Bardeen (left plot) and Hayward (right plot) BHs , assuming a values of $a_\star$ identical to \cite{Calza:2024fzo,Calza:2024xdh,Calza:2025whq}.}
    \label{spectra comparison}
\end{figure}
\noindent Fig.~(\ref{spectra comparison}) illustrates a direct comparison between the photon spectra obtained using $\tt BlackHawk$ and the results published in \cite{Calza:2024fzo,Calza:2024xdh,Calza:2025whq}. The superimposed plots reveal an excellent agreement: the spectra overlap to such an extent that they are visually indistinguishable.
 We point out that the tables of regular and BCL BHs are obtained by a modified version of {\tt GrayHawk}, the second version of which, {\tt GrayHawk v2} \cite{Calza:2026wuf}, is released together with {\tt BlackHawk v3.0}. To further validate the reliability of {\tt BlackHawk v3.0} and propose quantitative comparisons with third-party results, it is worth mentioning that {\tt BlackHawk v3.0} inherits the accuracy of the gray-body factors computed by {\tt GrayHawk} \cite{Calza:2025whq} and {\tt GrayHawk v2} \cite{Calza:2026wuf}, where many comparisons are considered and quantified using residuals.

\section{Conclusions}
\label{sec:conclusion}

In this work, we presented \texttt{BlackHawk v3.0}, a major update of the public code dedicated to the computation of Hawking radiation spectra from BHs. Building upon the numerical settings developed in the previous versions of the code, \texttt{BlackHawk v3.0} significantly enlarges the class of BH geometries that can be investigated within a unified and flexible framework, with a particular emphasis on regular and quantum-gravity-inspired black holes. The main novelty of this release is indeed the implementation of several new spherically symmetric metrics, including the Bardeen and Hayward regular black holes, the Simpson--Visser and Peltola--Kunstatter black-bounce geometries, the D'Ambrosio--Rovelli metric, and the Babichev--Charmousis--Leh\'ebel (BCL) black hole arising in Horndeski gravity. For all these metrics, we computed the corresponding Hawking temperatures and greybody factors, allowing the determination of primary Hawking emission spectra for particles of different spins. 

The computation of greybody factors has been performed using dedicated numerical routines based on the companion code \texttt{GrayHawk}, which solves the associated scattering problem for test fields propagating on generic BH backgrounds. These new implementations considerably extend the phenomenological reach of \texttt{BlackHawk}. In particular, they enable the investigation of how spacetime geometry modifications affect Hawking evaporation and, consequently, the observational constraints on primordial black holes (PBHs). Since both the Hawking temperature and greybody factors are sensitive to the underlying metric, regular and modified black holes can exhibit evaporation spectra that differ substantially from the Schwarzschild case. This may lead to important consequences for PBH dark matter constraints, cosmological observables, and the phenomenology of quantum-gravity-inspired compact objects. A key aspect of this release is the adoption of a conservative and model-independent strategy regarding the evolution of regular black holes. While instantaneous Hawking spectra are computed for all implemented metrics, we refrain from modeling the time evolution of regularizing parameters in the absence of a robust and universally accepted dynamical framework. This choice reflects the present theoretical uncertainties surrounding the microscopic interpretation of regular black holes and their evaporation dynamics. 

Beyond the implementation of new metrics, \texttt{BlackHawk v3.0} also includes several technical improvements and optimizations aimed at enhancing the flexibility, stability, and efficiency of the code. These updates facilitate the inclusion of additional metrics and particle sectors, making the framework suitable for future studies involving modified gravity, beyond-the-Standard-Model physics, and alternative quantum-gravity scenarios. The study of Hawking radiation remains one of the most promising avenues toward understanding the interplay between gravity, quantum mechanics, and particle physics. In this context, numerical tools such as \texttt{BlackHawk} play an essential role in connecting theoretical models with observational predictions. We hope that the new capabilities introduced in \texttt{BlackHawk v3.0} will provide a useful platform for future investigations of black hole evaporation, primordial black holes, and quantum-gravity-inspired phenomena.

Several possible extensions of the code are currently under consideration. These include the implementation of rotating regular black holes, dynamical evaporation models for regular metrics, additional modified-gravity solutions, more realistic electromagnetic charge effects, and improved treatments of beyond-the-Standard-Model particle emission. We also anticipate further developments toward a more systematic treatment of greybody factors for generic spacetimes and the inclusion of additional observational interfaces relevant for cosmology and astrophysics.

\begin{acknowledgments}

We want to thank Michael Zantedeschi for fruitful discussions. DP acknowledges support from the Istituto Nazionale di Fisica Nucleare (INFN) through the Commissione Scientifica Nazionale 4 (CSN4) Iniziativa Specifica “Quantum Fields in Gravity, Cosmology and BHs” (FLAG). This publication is based upon work from the COST Action CA21136 “Addressing observational tensions in cosmology with systematics and fundamental physics” (CosmoVerse), supported by COST (European Cooperation in
Science and Technology). 
\end{acknowledgments}

\section{Appendix}

\subsection{Mathematica notebooks derived from GrayHawk}

This part of the appendix contains a brief description of the scripts obtained by modifying the Mathematica code {\tt GrayHawk} and used to compute the tables in ``{\tt scr/tables/gamma\_tables...}" for the Hayward, Bardeen, SV, DR, PK, and BCL BHs. The scripts will be found in the folders: \\

{\tt >>> scripts/greybody\_scripts/greybody\_factors/GrayHawk/Bardeen}\\

{\tt >>> scripts/greybody\_scripts/greybody\_factors/GrayHawk/Hayward}\\

{\tt >>> scripts/greybody\_scripts/greybody\_factors/GrayHawk/Simpson-Visser}\\

{\tt >>> scripts/greybody\_scripts/greybody\_factors/GrayHawk/Peltola-Kunstatter}\\

{\tt >>> scripts/greybody\_scripts/greybody\_factors/GrayHawk/D'Ambrosio-Rovelli}\\

{\tt >>> scripts/greybody\_scripts/greybody\_factors/GrayHawk/BCL}\\
\\
Each of the folders described above contains four files, corresponding to the field spin values
\(s = 0, 1/2, 1, 2\), as well as an additional subfolder labelled ``{\tt For Fits}". The files
are self-contained code modules that compute the quantities \(Q^{s}_{l m}\) as defined in
Eq.~(\ref{Qfactors}). These files are based on ad hoc modifications of {\tt GrayHawk}, where the
introduction of nested {\tt for}-loops automates the evaluation of gray-body factors for multiple
values of the regularization parameter \(\mathfrak{l}\) and for different angular momentum \ l. The resulting contributions are summed to obtain \(Q^{s}_{l m}\), and the corresponding
data are organized into tables for various values of \(\mathfrak{l}\), which are exported in
{\tt .txt} format.

\noindent The {\tt For Fits} subfolder contains analogous scripts optimized to compute \(Q^{s}_{l m}\) at
energies outside the interval \(0.01 \leq x \equiv E r_S \leq 5\). Specifically, calculations are
performed at \(x = 0.001, 0.005, 25,\) and \(50\). These additional data points are used to
extrapolate the fitting functions beyond the range covered by the tabulated energies. The folder \\

{\tt >>> scripts/greybody\_scripts/greybody\_factors/GrayHawk/Formatter}\\

\noindent contains the file {\tt formatter.py} which is use to reformat of the .txt tables in such a way to obtain the one in reported in \\

{\tt >>> src/tables/gamma\_tables/...}\\

\subsection{Notebooks for the Kerr Case}

\noindent Here, we briefly describe the codes used to compute the gray-body factors for Kerr BHs in the direct method. The {\tt python} scripts and a {\tt README} are in the folder:\\

{\tt >>> scripts/greybody\_scripts/greybody\_factors/kerr}\\

\noindent The files contained in that folder are:
\begin{description}
    \item[Script containing main {\tt python} class] {\tt greybody\_Kerr.py}
    \item[File containing Kerr background class] {\tt background.py}
    \item[Script containing approximation for angular eigenvalues until order six, following Ref.~\cite{Dong:2015yjs}] {\tt lms.py}
    \item[Teukolsky equations and required defintions] {\tt teuk\_eqs.py}
    \item[Algorithm to obtain absorption probabilities] {\tt radial\_eqs.py}
    \item[Formatting] {\tt format.py}
    \item[Example script to run the code and obtain graybody tables] {\tt Glms.py}
\end{description}

The main class in {\tt greybody\_Kerr.py}, called {\tt Hawk\_spectrum}, requires as input the dimensionless parameter $w = GME$, the spin of the particle {\tt s}, and the background quantities for the Kerr spacetime, formatted according to the class {\tt Background} in {\tt background.py}. 
This main class contains the function {\tt d2N\_dEdt(l, m)} which computes the Hawking spectrum $Q_{slm}(\omega)$ according to Eq.~\eqref{Qfactors} after solving the radial part of the Teukolsky equation. 
Note that for the angular eigenvalues we have $_sQ_l^m = \ _s E^m_l -s(s+1) + a^2\omega^2-2a\omega m$, with$\ _s E^m_l$ given by an expansion up to six order in $a_\star GM \omega$, with coefficients as given in Ref.~\cite{Dong:2015yjs}.

We evolve the radial Teukolsky equation from a value of $x=10^{-4}$ up to $x=10^3$, assuming as initial condition the Frobenius series given in Eq.~\eqref{eq:fro_series}, with coefficients $a_n$ given in {\tt teuk\_eqs.py}. Then, we compare our numerical solution with the asymptotic form given in Eq.~\eqref{eq:asympt} to extract $R_{s,in}^{lm}(\omega)$, and then we compute $\Gamma_{lm}^s(\omega)$.
The algorithm for the evolution of the radial equations is contained in the {\tt radial\_eqs.py} files.

To obtain the Page factors $f,g$, one needs to use the function {\tt Ngf\_dEdt(l, m)} in the main {\tt Hawk\_spectrum} class for given values of $l,m$ quantum numbers.
An example of how to use these functions is provided in {\tt Glms.py}, which produces the Page factors in the format used by the main {\tt BlackHawk} code.

\clearpage
\newpage

\bibliographystyle{apsrev4-1}
\bibliography{main.bib}

\end{document}